\documentclass[12pt,onecolumn,preprint]{elsarticle}
\begin{document}
\begin{frontmatter}
\title{A Hamiltonian for 1/1 Rotational Secondary Resonances, and Application to Small Satellites of Saturn and Jupiter}

\author{N. Callegari Jr.\corref{cor1}}
\ead{nelson.callegari@unesp.br}
\address{Instituto de Geoci\^{e}ncias e Ci\^{e}ncias Exatas, Unesp - Univ Estadual Paulista,}
\address{Departamento de Estat\'{\i}stica, Matem\'{a}tica Aplicada e Computa\c {c}\~{a}o.}
\address{Av. 24-A, Rio Claro/SP/Brazil, CEP 13506-900.}

\cortext[cor1]{Corresponding author}



\begin{abstract}

In this work, we study the dynamics of rotation of the small satellites Methone and Aegaeon and revisit previous works on the rotation of Prometheus, Metis, and Amalthea. In all cases, the surfaces of section computed with the standard spin-orbit model reveal that the synchronous regime with small amplitude of libration shares another large domain in the phase space. We reproduce and apply the hamiltonian theory given in Wisdom (2004) to analytically characterize the detected structure as being a secondary resonance where the period of oscillation of the physical libration is similar to the orbital period of the satellite. We also show that the amplitude of libration around the secondary resonance is always larger than in the case of the other mode. Since the current rotational states of these sorts of satellites should be synchronous, our results can be considered in evolutionary studies of their rotation.

\end{abstract}

\begin{keyword}
Aegaeon; Amalthea; Metis; Methone; Prometheus; Saturnian and Jupiter small satellites; Spin-Orbit resonance.
\end{keyword}
\end{frontmatter}

\section{Introduction}

Consider the rotation of a rigid and homogeneous satellite orbiting a punctual planet in the case the satellite rotates around a single axis corresponding to its largest moment of inertia. The mutual perturbations between the bodies belonging to multiple systems or due to the non-sphericity of the secondary are neglected in this work, such that the motion of the system is governed by the laws of the two-body problem (Goldreich and Peale 1966).

A hamiltonian for the sort of problem is given by:

\begin{equation}
H(\theta,p,t)=\frac{p^2}{2C}-\frac{\epsilon^2n^2C}{4}\left[\frac{\textsf{a}}{r(t)}\right]^3
\cos{2[\theta-f(t)]},\label{1}
\end{equation}
where $\theta$ is an angle of rotation of the satellite measured from an inertial line; $p=C\dot\theta$ is the angular momentum of rotation around the $z-$axis assumed to be perpendicular to the orbital plane, and $\dot\theta$ is the angular velocity of rotation; $\epsilon^2=3\frac{B-A}{C}$, where $A<B<C$ are the moments of inertia around the principal axes of the satellite $x$, $y$, $z$, respectively; $\textsf{a}$, $e$, $f(t)$, $r(t)=\frac{\textsf{a}(1-e^2)}{1+e\cos f(t)}$, $n$ are the semi-major axis, orbital eccentricity, true anomaly, planet-satellite distance, and the mean motion of the satellite, respectively.

It is well-known that the mid-sized satellites of the outer planets and the Moon rotate synchronously such that, roughly speaking, the velocity of angular rotation is equal to the mean motion (see Goldreich and Peale 1966, Peale 1977, Melnikov and Shevchenko 2022). The synchronous orbit-rotation resonance is characterized by the oscillation around zero of the angle $\Psi(t)=\theta(t)-f(t)$ present in Equation (\ref{1}). $\Psi$ contains all main components of the spin-orbit perturbations, namely, the optical, free, and forced librations (see Murray and Dermott 1999, Callegari and Ribeiro 2015, and references therein). In the case of a small amplitude of libration of $\Psi$ and small orbital eccentricity of the satellite, the physics of the synchronous resonance can be interpreted by an analog of the forced harmonic oscillator (see Murray and Dermott 1999). Suppose also to model the shape of the satellite with a Roche ellipsoid with semi-axes $a$, $b$, and $c$ in the $x$, $y$, $z$ directions, respectively; in this case, $a>b>c$ (see Callegari and Rodr\'{i}guez 2013, and references therein). Denote the free frequency by $\omega$ (corresponding to the frequency in the harmonic approximation). The average linear theory\footnote{W.r.t. mean anomaly and valid for small eccentricity.} of synchronous resonance predicts that the ratio of the free and orbital frequencies $\omega/n$ is given by:

\begin{eqnarray}
\epsilon\equiv\frac{\omega}{n}=\left(3\frac{B-A}{C}\right)^{1/2}=\left[3\left(\frac{a^2-b^2}{a^2+b^2}\right)\right]^{1/2}. \label{2}
\end{eqnarray}
Thus, when $\epsilon$ is a rational number, an inner resonance can rise due to the commensurability between $n$ and $\omega$. We have therefore a definition for secondary resonance within the synchronous regime (in the case of a synchronous rotating satellite). The approximation of a simple oscillator is no longer enough to describe the rotational dynamics in this situation, and additional analyses (e.g. perturbation theory) are necessary to describe the dynamics of rotation in more detail.

The case of the Saturnian satellite Enceladus studied in Wisdom (2004) is a good example to illustrate the dynamics of secondary resonances. Under the hypothesis of a homogeneous body, and based on \emph{Voyager} spacecraft data, the dimensions of Enceladus were $a=256.3\pm0.3$ km, $b=247.3\pm0.3$ km, $c=244.6\pm0.3$ km (Dermott and Thomas 1994). From Equation (\ref{2}) we obtain $\epsilon\sim0.327$, a value close to $\frac{1}{3}$, so that Wisdom conjectured if Enceladus would be currently rotating close to the $\frac{\omega}{n}=\frac{1}{3}$ secondary resonance\footnote{After\emph{ Cassini} images, the amplitude of the physical libration of Enceladus has been detected ($\sim0.12$ degree; Thomas et al. 2016). This value is too high and therefore is not consistent with a homogeneous satellite. The best fitting given in Thomas et al. (2016) corresponds to a multi-layered satellite including a global subsurface ocean. They also determine $a=256.2\pm0.3$ km, $b=248.6\pm0.2$ km, $c=252.24\pm0.2$ km leading to $\epsilon\sim0.300$.}.

While it is certain that current rotational states of \emph{all} mid-sized satellites of the giant planets are synchronous, the same is not necessarily true for \emph{all} other smaller secondary companions having mean radius of the order of a few dozens kilometer. Several of them are irregularly shaped, and due to this, $\epsilon$ is large. Let us consider the case of Amalthea (J15), a member of the inner group of Jupiter satellites composed of Metis, Adrastea, Amalthea, and Thebe. Amalthea is significantly out of round with dimensions $a \times b \times c$ of the order of $\sim$ $125 \times \sim 73 \times \sim64$ km such that $\epsilon\sim1.214$. These values are given after analyses of \emph{Galileo} tour in Jupiter system (Thomas et al. 1998; see also Pashkevich et al. 2021, and references therein). The phase space of rotation of Amalthea shows the existence of two resonant modes within the synchronism denoted by ``$\alpha$-resonance'' and ``$\beta$-resonance''; this property of the synchronous regime has been named by the ``Amalthea effect'' (e.g. Melnikov and Shevchenko 2008). The existence of such kind of ``bifurcation'' inside the synchronous domain leads to questioning and discussions on topics related to the evolution of rotation and the true current equilibrium configuration of the satellite (see discussions in Melnikov and Shevchenko 2002, and references therein). Another body with similar rotational properties of the phase space is the Prometheus (Melnikov and Shevchenko 2008), from Saturn, a close-in satellite with a mean radius of 42.8 km (see Table 1).

\begin{table}
 \center
 \begin{minipage}{200mm}
  \caption{Selected physical and orbital parameters of the satellites studied in this work. In the case of Saturn, \\the values of the orbital eccentricities correspond to the maximum geometric values (Callegari et al. 2021;\\ Ceccatto et al. 2021, Callegari and Rodríguez 2023).}
   \vspace{0.5cm}

\begin{tabular}{cccccc}
\hline
                    &Methone$^a$             & Aegaeon$^a$      & Prometheus$^a$ &  Amalthea$^b$ & Metis$^b$       \\

\hline \hline

semi-major axis (km)&   194,707           & 168,031     & 140,029    &  181,400   &  128,000       \\ \hline

eccentricity        &   0.0014            & 0.0005      & 0.00226    &  0.0031    &  0.0002            \\ \hline

dimensions (km)     & $1.94 \times$  &     $0.7 \times$ & $68.5 \times$        &  $125 \times$    & $30 \times$            \\
                    &  $1.29 \times$ &    $0.25 \times $   & $ 40.5 \times$    &  $73 \times $   & $20 \times $            \\
                    &  1.21         &       0.20       &  28.1                &    64             & 17            \\\hline

$\epsilon$          & 1.077              &  1.5236      &  1.1717    &  1.2140      &  1.074       \\ \hline

\end{tabular}

{$(^a)$ Thomas and Helfenstein 2020;}
{$(^b)$ Thomas et al. 1998; Pashkevich et al. 2021}

\end{minipage}
\end{table}


The objects we study in this project belong to another class of even smaller inner satellites of the Jovian planets having mean diameters of a few kilometers or even less. We consider in this work the Saturnian satellites Methone (S/2004 S 1) and Aegaeon (S/2008 S 1) discovered by the \emph{Cassini} spacecraft in 2004 (Porco 2004) and 2008 (Porco 2009), respectively. Similarly to Amalthea and Prometheus, their phase spaces also display the co-existence of the $\alpha$ and $\beta$-``resonances'', as we will show. Methone, Aegaeon (and Pallene) are ellipsoidal-like satellites\footnote{The other denominations are irregular (e.g. Prometeus, Pandora, Epimetheus, Janus, Telesto, Calypso, and Helene), and irregular with equatorial ridges (e.g. Atlas, Pan, Daphnis) (Thomas and Helfenstein 2020).}. We will focus next and throughout the paper on the case of Methone, and at the end of the paper, similar results will be shown and discussed for Aegaeon and Prometheus. The cases of Amalthea and Metis will also be revisited given the theory developed here.

To investigate the rotation of Methone, we apply the standard model of spin-orbit resonance. We utilize the Everhart (1985) algorithm to solve numerically the differential equations derived from hamiltonian (\ref{1}). The full equation is non-autonomous such that, at first, we compute the surfaces of section (hereafter denoted by SOS), a practical procedure often applied in studies of rotation after Wisdom et al. (1984). In this technique, the pair of generalized variables $(\theta,\dot\theta/n)$ is plotted every time the satellite passes through the pericenter of its orbit. Fig. \ref{1}a) shows the SOS for Methone in the vicinity of the synchronous domain. The fixed points of the $\alpha$ and $\beta$-``resonances'' are indicated by vertical arrows. Note that the regimes are separated by a new separatrix (red curve in Fig. \ref{1}a)), while they are encompassed by a thin chaotic layer associated to the synchronous domain.

$\alpha$ and $\beta$-regimes share the synchronous domain, so it is worth giving a cinematic description of the rotation of clones of Methone departing from initial conditions very close to their fixed points to identify their differences. Figs. \ref{1}(b,c) show the time variations of $\frac{\dot\theta}{n}$ and $\Psi=\theta-f$ corresponding to the $\alpha$ and $\beta$-regimes, respectively. The initial conditions are $\theta_0=0$ in both cases, and $\frac{\dot\theta(0)}{n}=1.0211$, $\frac{\dot\theta(0)}{n}=0.45$ in b) and c), respectively.

In the case of the $\beta$-regime (Fig. \ref{1}b)), $\Psi$ oscillates around zero with amplitude of $\sim1$ degree and $\frac{\dot\theta}{n}$ oscillates with amplitude of $\sim0.02$.
In the case of the $\alpha$-regime (Fig. \ref{1}c)), $\Psi$ oscillates around zero with amplitude of $\sim 30$ degree. $\frac{\dot\theta}{n}$ oscillates around the unit with amplitude of $\sim0.5$ such that at the minima, the satellite is always located at the pericenter of its orbit (full circles in Fig. \ref{1}c), bottom). Thus, due to this relatively large amplitude of oscillation of $\frac{\dot\theta}{n}$, the SOS displays $\frac{\dot\theta}{n}\sim 0.45$, the same as that of the initial value.


\newpage
\begin{figure}[h]
\centerline{\includegraphics[width=36pc]{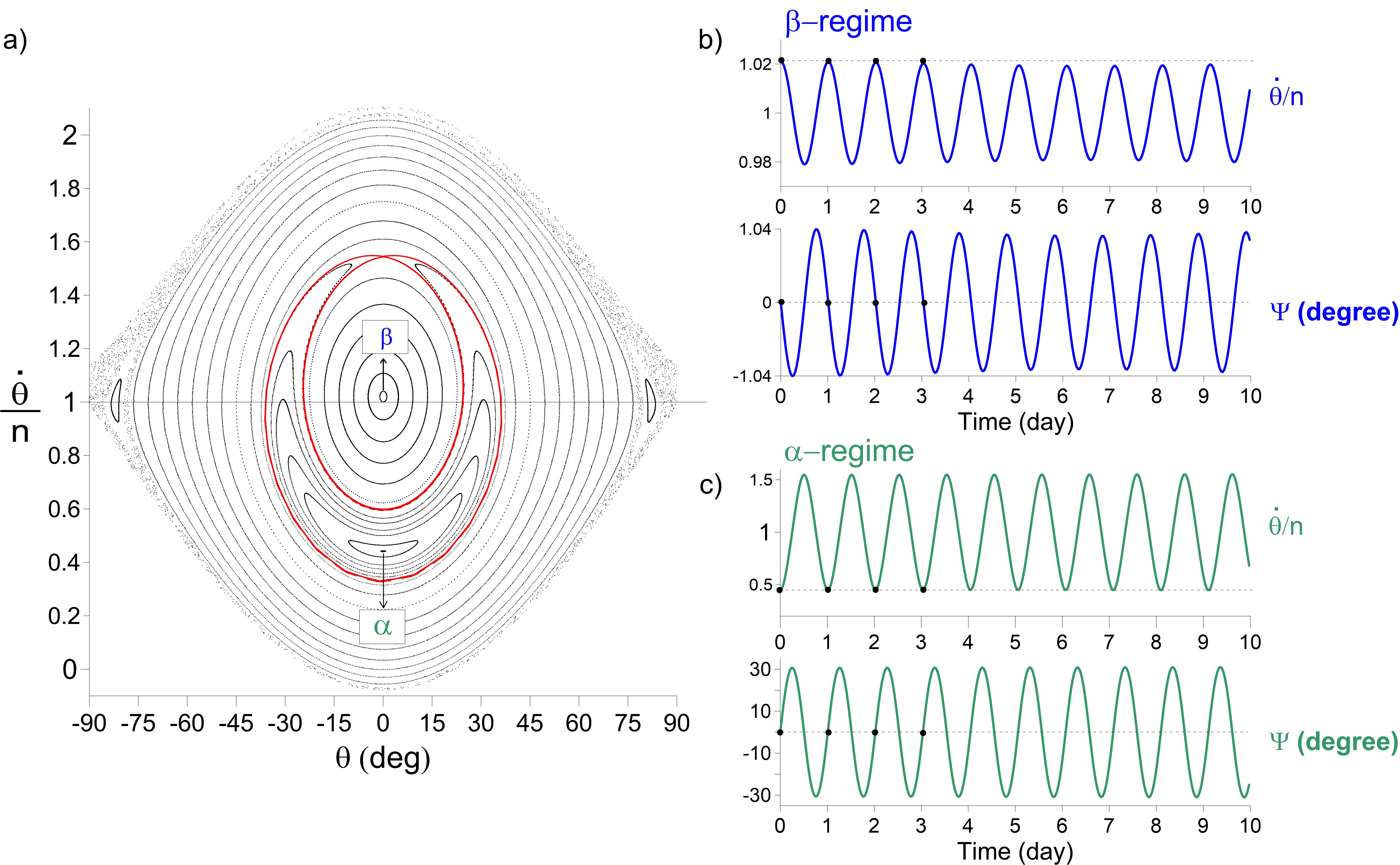}}
\caption{(\textbf{a)} Surfaces of section of solutions of the model (\ref{1}) applied to the satellite Methone. Each orbit has been
integrated for 500 orbital periods of the satellite. The sections (projections) are made at each orbital revolution, and all the time the true anomaly of the satellite reaches $f=0$. The $x$ and $y$-axes are the (projected) rotation angle ($\theta$) around the axis of the largest moment of inertia and the ratio of the angular velocity of rotation and the mean motion, $\frac{\dot\theta}{n}$. The vertical arrows indicate the fixed points associated with the $\beta$ and $\alpha$-regimes of rotation. The horizontal line  gives the straight $\frac{\dot\theta(t)}{n}=1$. The numerical values of the parameters of the model adopted in this work which are necessary in the numerical solution are: $\textsf{a}=194,707.6$ km, $e=0.0014$, $a=1.94$ km, $b=1.29$ km (see Section 1 for definitions). \textbf{(b,c)} Time variations of $\dot\theta(t)/n$, $\Psi(t)=\theta(t)-f(t)$ for two trajectories within the domains of the $\beta$ and $\alpha$-regimes, respectively. $\theta_0=0$ in both cases and $\frac{\dot\theta(0)}{n}=1.0211$, $\frac{\dot\theta(0)}{n}=0.45$ in b) and c), respectively. The horizontal dashed lines plotted in (b,c) guide the solutions close to the fixed points at the pericenter passage (indicated by full circles).}
\label{<1>}
\end{figure}

\newpage

Since the Methone's phase space of rotation displays additional structures within the synchronous regime, all discussion given above leads us to investigate the existence of secondary resonances. According to the most recent analyses of data taken from \emph{Cassini} measurements, Thomas and Helfenstein (2020) determined that the dimensions of Methone are $a=1.94\pm0.02$ km, $b=1.29\pm0.04$ km, $c=1.21\pm0.02$ km so that $\epsilon\sim1.077$. This value of $\epsilon$ is close to the unit so that we conjecture if a 1/1 secondary resonance of the type $\frac{\omega}{n}\sim\frac{1}{1}$ would explain the existence of $\alpha$ and $\beta$-structures within the synchronous domain. To prove the conjecture we return to the Wisdom's (2004) paper. He developed a simple analytical model for spin-orbit problem in hamiltonian formalism aiming to explain the phase space of Enceladus\emph{ in the close vicinity} of the synchronous regime. Having in hand an expanded version of hamiltonian of Equation (\ref{1}), he selected the correct terms and explained analytically the properties of the secondary 1/3 resonance. We follow his steps and reproduce its main results, and we choose those terms in hamiltonian which are proportional to the arguments containing the 1/1 secondary resonance. By adopting a similar methodology given in Wisdom (2004), we successfully explain the co-existence of the $\alpha$ and $\beta$-regimes in synchronous resonance by comparing the SOS with level curves of the constructed hamiltonian\footnote{It is worth noting that the mapping of the 1/1 secondary resonance within the synchronous regime has been done in many works where very sophisticated models have been adopted (see Gkolias et al. 2016, Gkolias et al. 2019, Lei 2023.) }.

The presentation of the results of this paper is divided as follows. In Section 2, we deduce the expanded hamiltonian developed in Wisdom (2004) in detail, showing the terms of the hamiltonian explicitly (that ones related to equation 20 of the referred paper); some extensions of Wisdom's model are also given. In Sections 3.1 and 3.2, we show exactly how the perturbed hamiltonian can give rise to the secondary resonances and write the hamiltonian for the 1/1 secondary resonance. The level curves applied in the case of Methone are shown in Section 3.3. Analytical estimation of the bifurcations of the level curves are provided in Section 3.4. In Section 3.5, we provide an equation to obtain the amplitude of oscillation at the fixed points of the $\alpha$ and $\beta$-regimes as a function of $\epsilon$; we show that the amplitude of oscillation around the secondary resonance is always larger than the case of the $\beta$-resonance. In Section 4, we analyze the cases of Aegaeon and Prometheus,  and Section 5 is devoted to conclusions and more general discussion involving other satellites. Since the phase space of rotation of Methone, Aegaeon, Prometheus, Amalthea, Metis, and others share similar structures and properties within the synchronous regime, we are led to question their current states of rotation.


\section{Expansion of the hamiltonian (\ref{1})}

\subsection{An integrable hamiltonian}


The hamiltonian (\ref{1}) is analogous to that one of a simple \emph{non-autonomous} pendulum. An expanded version of the hamiltonian can be obtained resulting in a sum of autonomous-like pendulums. This can be easily achieved after substituting in hamiltonian (\ref{1}) the developed forms of $\left[\frac{\textsf{a}}{r(t)}\right]^3$, $\cos{f(t)}$, $\sin{f(t)}$:

\begin{eqnarray}
\left[\frac{\textsf{a}}{r(t)}\right]^3\approx1+3e\cos{nt}+\frac{3}{2}e^2(1+3\cos{2nt}),\nonumber \\
\cos{f(t)}\approx\cos{nt}+e(\cos{2nt}-1)+\frac{9}{8}e^2(\cos{3nt}-\cos{nt}),\label{3}\\
\sin{f(t)}\approx\sin{nt}+e\sin{2nt}+\frac{1}{8}e^2(9\sin{3nt}-7\sin{nt}).\nonumber
\end{eqnarray}

We have:

\begin{eqnarray}
H(\theta,p,t)=\frac{p^2}{2C}
&-&\frac{\epsilon^2n^2C}{4}\left(+\frac{17}{2}e^2\right)\cos{(2\theta-4nt)}\nonumber\\
&-&\frac{\epsilon^2n^2C}{4}\left(+\frac{7}{2}e\right)\cos{(2\theta-3nt)}\nonumber\\
&-&\frac{\epsilon^2n^2C}{4}\left(1-\frac{5}{2}e^2\right)\cos{2(\theta-nt)}\nonumber\\
&-&\frac{\epsilon^2n^2C}{4}\left(-\frac{1}{2}e\right)\cos{(2\theta-nt)}+\ldots .\label{4}
\end{eqnarray}
Each term in the right side of (\ref{4}) proportional to $\epsilon^2$ can be considered a distinct disturbing resonance associated to the commensurabilities  $\dot\theta=2n$, $\dot\theta=\frac{3}{2}n$, $\dot\theta=n$ and $\dot\theta=\frac{n}{2}$, defining therefore the 2:1, 3:2, 1:1 (synchronous), and the 1:2 spin-orbit resonant states.


We can isolate a single resonance after applying the average principle (see footnote 1) and we can note that, whatever resonance, the resulting hamiltonian is always analogous to a simple \emph{non-autonomous} pendulum. Canonical transformations can make each of the resonances given in Equation (\ref{4}) locally integrable. For instance, consider the synchronous resonance and define $H_0$ by:

\begin{eqnarray}
H_0(\theta,p,t)=\frac{p^2}{2C}-\frac{\epsilon^2n^2C}{4}\left(1-\frac{5}{2}e^2\right)\cos{2(\theta-nt)}.
\label{5}
\end{eqnarray}
Given the canonical transformation:

\begin{eqnarray}
        \phi&=&\theta-nt \label{6} \\
        \Phi&=&p-nC\nonumber,
\end{eqnarray}
the new $H'_0$ is given by:

\begin{eqnarray}
H'_0(\phi,\Phi,-)&=&-n\Phi+\frac{(\Phi+nC)^2}{2C}-\frac{\epsilon^2n^2C}{4}\cos{2\phi}\Rightarrow\nonumber\\
          &=&\frac{\Phi^2}{2C}-\frac{\epsilon^2n^2C}{4}\cos{2\phi},\label{7}
\end{eqnarray}
where the term $-n\Phi$ comes from phase space extension (see Sussman and Wisdom 2001, Ferraz-Mello 2007). We have also neglected the term in $5e^2/2$ since the orbital eccentricity of Methone is of the order of $10^{-3}$. Moreover, an irrelevant additive constant ($n^2C/2$) is not included in (\ref{7}).

Thus, we have an \emph{integrable} hamiltonian for the synchronous spin-orbit resonance. Since the hamiltonian (\ref{7}) is analogous to the simple autonomous pendulum, we can use the equations of transformation to the action-angle variables developed to the simple pendulum which are given in several books. In this work, we consider only the libratory regime in the case of small oscillations.

The action variable $J$ is defined as being the area on the plane $(\phi,\Phi)$ divided by $2\pi$. The conjugated action $\psi$ varies linearly in time. The new hamiltonian depends only on the action and can be written as follows:

\begin{eqnarray}
H^{''}_0(-,J)=-\frac{\epsilon^2n^2C}{4}+\epsilon nJ-\frac{1}{4C}J^2-\frac{1}{16\epsilon nC^2}J^3+\ldots.\label{10}
\end{eqnarray}

The algebraic expression between the new variables $(\psi,J)$ and the old ones $(\phi,\Phi)$ is:
\begin{eqnarray}
\sin{2\phi}\approx2\phi=F_1\sin{\psi}+F_3\sin{3\psi}+\ldots,\label{11}
\end{eqnarray}
where
\begin{eqnarray}
F_1&\approx&\frac{2\sqrt{2}}{\lambda^{1/2}}J^{1/2}+\frac{\sqrt{2}}{2\lambda^{3/2}}J^{3/2}+\frac{ 23\sqrt{2} }{ 64\lambda^{5/2}}J^{5/2},\label{12} \\
F_3&\approx&\frac{\sqrt{2}}{12\lambda^{3/2}}J^{3/2}+\frac{3\sqrt{2}}{ 32\lambda^{5/2} }J^{5/2},\label{13}
\end{eqnarray}
where $\lambda=\epsilon nC$.

Equations (\ref{10})-(\ref{13}) can be obtained from Wisdom (2004). Their deduction results of applying the pendulum model to the hamiltonian (\ref{7}) in the case of small oscillation approximation (see for instance appendix B.3 in Ferraz-Mello 2007).
\hspace{1cm}

\subsection{Perturbation of the integrable hamiltonian}

Let us consider now the time-dependent perturbations on the integrable part of the synchronous resonance of the hamiltonian such that

\begin{eqnarray}
H'(\phi,\Phi,t)=H'_0+H'_1, \label{14}
\end{eqnarray}
where $H'_0$ is given in (\ref{7}) and $H'_1$ is the perturbation. We will include in $H'_1$ the two neighboring resonances of the synchronous resonances, namely, 3:2 and 1:2, which are proportional to $e$ (see Equation (\ref{4})). In terms of old variables $(\phi,\Phi)$ we can obtain from (\ref{4}):

\begin{eqnarray}
H'_1(\phi,\Phi,t)=-\frac{\epsilon^2n^2C}{4}
\left[\frac{7e}{2}\cos{(2\phi-nt)}-\frac{e}{2}\cos{(2\phi+nt)}\right].\label{15}
\end{eqnarray}

To write $H'_1$ in terms of the action-angle variables we must consider the transformation equation (\ref{11}). The algebraic manipulations can be much more simple considering the limit of small oscillations (Wisdom 2004). Thus, first write (\ref{15}) as follows:

\begin{eqnarray}
H'_1(\phi,\Phi,t)&=&-\frac{\epsilon^2n^2C}{4}
\left[\frac{7e}{2}(C_2\cos{nt}+S_2\sin{nt})-\frac{e}{2}(C_2\cos{nt}-S_2\sin{nt})\right], \label{16}\\
C_2&=&\cos{2\phi}\approx1-\frac{1}{2}(2\phi)^2, S_2=\sin{2\phi}\approx2\phi-\frac{1}{6}(2\phi)^3. \label{17}
\end{eqnarray}

After substituting the expression (\ref{11}) in (\ref{17}) and the result in (\ref{16}), we can obtain the perturbation $H''_1(\psi,J,t)$ as follows\footnote{The addition of the next term in the expansion of Equation (\ref{11}) does not generate additional terms of order of $J^3$  in the cases of the resonances 3/1 and 1/1 considered in this work. The same is not necessarily true in other cases, like the 5/1 and 6/1 secondary resonances.}:
\begin{eqnarray}
H''_1(\psi,J,t)&=&\epsilon^2n^2Ce\times\nonumber\\
&&[(3/16)F_1^2 +(3/16)F_3^2-3/4]\cos{nt}+\nonumber\\
&&[(-1/2)F_1+(1/16)F_1^3-(1/16)F_1^2F_3+(1/8)F_1F_3^2]\cos{(nt-\psi)}+\nonumber\\
&&[(-3/32)F_1^2+(3/16)F_1F_3]\cos{(nt-2\psi)}+\nonumber\\
&&[(1/16)F_3^3-(1/48)F_1^3+(1/8)F_1^2F_3-(1/2)F_3]\cos{(nt-3\psi)}+\nonumber\\
&&((-3/16)F_1F_3\cos{(nt-4\psi)}+\nonumber\\
&&[(1/16)(F_1F_3^2-F_1^2F_3)]\cos{(nt-5\psi)}+\nonumber\\
&&(-3/32)F_3^2\cos{(nt-6\psi)}+\nonumber\\
&&(-1/16)F_1F_3^2\cos{(nt-7\psi)}+\nonumber\\
&&(-1/48)F_3^3\cos{(nt-9\psi)}+\nonumber\\
&&[(-1/16)F_1^3+(1/2)F_1-(1/8)F_1F_3^2+(1/16)F_1^2 F_3]\cos{(nt+\psi)}+\nonumber\\
&&[(3/16)F_1F_3-(3/32)F_1^2]\cos{(nt+2\psi)}+\nonumber\\
&&[(-1/16)F_3^3+(1/48)F_1^3+(1/2)F_3-(1/8)F_1^2F_3]\cos{(nt+3\psi)}+\nonumber\\
&&(-3/16)F_1F_3\cos{(nt+4\psi)}+\nonumber\\
&&[(1/16)(F_1^2F_3-F_1F_3^2)]\cos{(nt+5\psi)}+\nonumber\\
&&(-3/32)F_3^2\cos{(nt+6\psi)}+\nonumber\\
&&(1/16)F_1F_3^2\cos{(nt+7\psi)}+\nonumber\\
&&(1/48)F_3^3\cos{(nt+9\psi)}
\label{18}.
\end{eqnarray}

\section{Perturbed hamiltonian and the rise of secondary resonances}

\subsection{The 1/1 secondary resonance}

The transformation variable given in Equation (\ref{6}), $\phi=\theta-nt$, consists
of an approximation of the physical libration $\Psi$ valid for small eccentricity.

For small amplitude of oscillation, the frequency of the angle variable $\psi$ is the frequency of $\phi$ divided by $2\pi$. Thus, Equation (\ref{18}) can be utilized to study several secondary resonances. Inspection of (\ref{18}) shows the terms in arguments of the cosines
of the form $nt-k\psi$, where $k=1,2,\ldots$. Therefore, when $k$ is a multiple of $n$, we should have a secondary resonance.

For instance, after collecting the terms proportional to $\cos{(nt-3\psi)}$,

\begin{equation}
H''_1(\psi,J,t)=-\frac{\epsilon^2n^2Ce}{4}\left( \frac{1}{12}F_1^3+2F_3-\frac{1}{4}F_3^3-\frac{1}{2}F_1^2F_3  \right)\cos{(3\psi-nt)},
\label{19}
\end{equation}
we can obtain the disturbing part of the hamiltonian associated with the 3/1 secondary resonance utilized in the theory developed by Wisdom (2004) to study the rotation of Enceladus (see Section 1)\footnote{The last two terms in Equation (\ref{19}) are higher order and don't contribute to Wisdom's model of the 3/1 secondary resonance, which is valid up to $J^2$.}.


The problem we are considering in this work is related to the nature of the $\alpha$ and $\beta$-regimes of motion in the phase space of rotation of Methone and other satellites; as pointed out in Section 1, we conjectured that the coexistence of the regimes is explained by a 1/1 secondary resonance. We will prove our conjecture in the next two subsections after considering the terms factored by $\cos{(\psi-nt)}$ in (\ref{18}) so that we have the following disturbing part of the hamiltonian:

\begin{equation}
H''_1(\psi,J,t)=-\frac{\epsilon^2n^2Ce}{4}\left( +\frac{1}{4}F_1^2F_3+2F_1+\frac{1}{4}F_1^3-\frac{1}{2}F_1F_3^2  \right)\cos{(\psi-nt)}.
\label{20}
\end{equation}

From Equations (\ref{12}) and (\ref{13}), keeping the terms up to $O(J^2)$, after neglecting the constant terms we obtain:\\

\begin{equation}
H''_1(\psi,J,t)=e \left[ +\frac{3}{8}\sqrt{\frac{\epsilon n}{C}}\left(\sqrt{2J}\right)^3-\sqrt{\frac{\epsilon^3n^3}{C}}\sqrt{2J}  \right]\cos{(\psi-nt)}.\label{319}
\end{equation}

\subsection{The second extension of the phase space and the non-singular variables}

Including terms up to $J^2$ in Equation (\ref{10}) and discarding constants terms we have

\begin{eqnarray}
H^{''}_0(-,J)= \epsilon nJ-\frac{1}{4C}J^2.
\label{318}
\end{eqnarray}

Consider the canonical transformation

\begin{eqnarray}
        \psi'&=&\psi-nt, \nonumber\\
        J'&=&J. \label{21}
\end{eqnarray}

From (\ref{318}) and (\ref{319}) the full hamiltonian  $H''_{1/1}(\psi',J',t)=H^{''}_0(J')+H''_1(\psi',J',t)$  becomes\\

\begin{eqnarray}
H'''_{1/1}(\psi',\hat{J})=-nJ'+\epsilon nJ'-\frac{1}{4C}J'^2+e \left[ +\frac{3}{8}\sqrt{\frac{\epsilon n}{C}}\left(\sqrt{2J'}\right)^3-\sqrt{\frac{\epsilon^3n^3}{C}}\sqrt{2J'}  \right]\cos{\psi'},\nonumber
\end{eqnarray}
where the term $-nJ'$ comes from a phase space extension.\\

Define

\begin{eqnarray}
\hat{J}&=&\frac{J'}{nC},\nonumber\\
\delta&=&\epsilon-1.\label{22}
\end{eqnarray}
where $\hat{J}$ is a dimensionless quantity (Wisdom 2004), and $\delta$ can be interpreted as being the distance of the exact resonance $1/1$ secondary resonance.\\

Factoring the parameter $n^2C$, we can obtain after trivial algebra the full hamiltonian in action-angle variables:

\begin{eqnarray}
H'''_{1/1}(\psi',\hat{J})=n^2C \left[\delta\hat{J}-\frac{\hat{J}^2}{4}+e\left(\frac{3}{8}\sqrt{\epsilon}
\left(\sqrt{2\hat{J}}\right)^3-\sqrt{\epsilon^3}\sqrt{2\hat{J}}\right)\cos{\psi'}  \right].
\label{23}
\end{eqnarray}\\

Define

\begin{equation}
H\equiv\frac{H'''_{1/1}}{n^2C}.
\label{24}
\end{equation}
It is straightforward to prove that $H$ is dimensionless so that the $H$ depends only on two parameters: $\epsilon$ and the orbital eccentricity $e$.

Let us consider the canonical transformation:
\begin{eqnarray}
        x&=&\sqrt{2\hat{J}}\cos{\psi'}\nonumber  \\
        y&=&\sqrt{2\hat{J}}\sin{\psi'}. \label{25}
\end{eqnarray}

Thus, the hamiltonian (\ref{24}) becomes:
\begin{equation}
H(x,y)=\frac{\delta}{2}(x^2+y^2)-\frac{1}{16}(x^2+y^2)^2+\frac{3}{8}e\sqrt{\epsilon}(x^2+y^2)x-e\sqrt{\epsilon^3}x.\label{26}
\end{equation}

\subsection{Atlas of the phase space}

Keeping in hand a time-independent one-degree-of-freedom hamiltonian for our problem, we can now explore the rotation phase space by computing the level curves of Equation (\ref{26}). We also aim to compare the level curves with the surfaces of sections and a suitable plane of variables which can be the same utilized in Fig. 1, $(\theta,\frac{\dot\theta}{n})$. To relate them to the variables $(x,y)$ (Equation (\ref{25})), we first recall that in the linear approximation of harmonic oscillator the relation between variables $(\phi,\Phi)$ (Equation (\ref{6})) and the angle-action $(\psi,J)$ are given by the canonical transformation
\begin{eqnarray}
\phi&=&\left(\frac{2J}{\lambda}\right)^{1/2}\sin{\psi},\nonumber\\
\Phi&=&(2J\lambda)^{1/2}\cos{\psi}, \label{27}
\end{eqnarray}
where $\lambda=\epsilon nC$. By fixing a instant of time ($t=0$ for instance), manipulating the Equations (\ref{6}), (\ref{25}), (\ref{27}), and by noting that at $t=0$, $\psi'=\psi$ (Equation (\ref{21})), we can show that,
\begin{eqnarray}
\theta&=&\frac{y}{\sqrt{\epsilon}},\nonumber\\
\frac{\dot\theta}{n}&=&x\sqrt{\epsilon}+1. \label{28}
\end{eqnarray}

Fig. \ref{2} shows the level curves of the hamiltonian (\ref{26}) for six values of $\epsilon$. Let $\epsilon_c$ be a critical value of  $\epsilon$. For $\epsilon<\epsilon_c$, only the equilibrium point associated with the $\alpha$-regime exists. The rising of the $\beta$-regime occurs at values of $\epsilon$ slightly larger than $\epsilon_c$, and now three equilibria centers share the $y$-axis: two of them are stable points, namely those associated to $\alpha$ and $\beta$-regimes, and the third one is an unstable equilibrium point associated to the separatrix of the $\beta$-regime. The plot in right-bottom panel in Fig. \ref{2} reproduces with good agreement the surface of section given in Fig. \ref{1}. The two curves given in blue and purple superposed to the level curves in the plot $\epsilon=1.077$ are the surfaces of section. The agreement is good, in spite small of differences in the position of the fixed points. The model can be improved by including high-order terms in the expansions in action to obtain better agreements (see Section 4).

Some investigations on the dependence of eccentricity of Methone on the phase space is given in Appendix.

Next we will study in detail the bifurcation of the regimes in the phase space and determine $\epsilon_c$ numerically.

\begin{figure}
\centerline{\includegraphics[width=42pc]{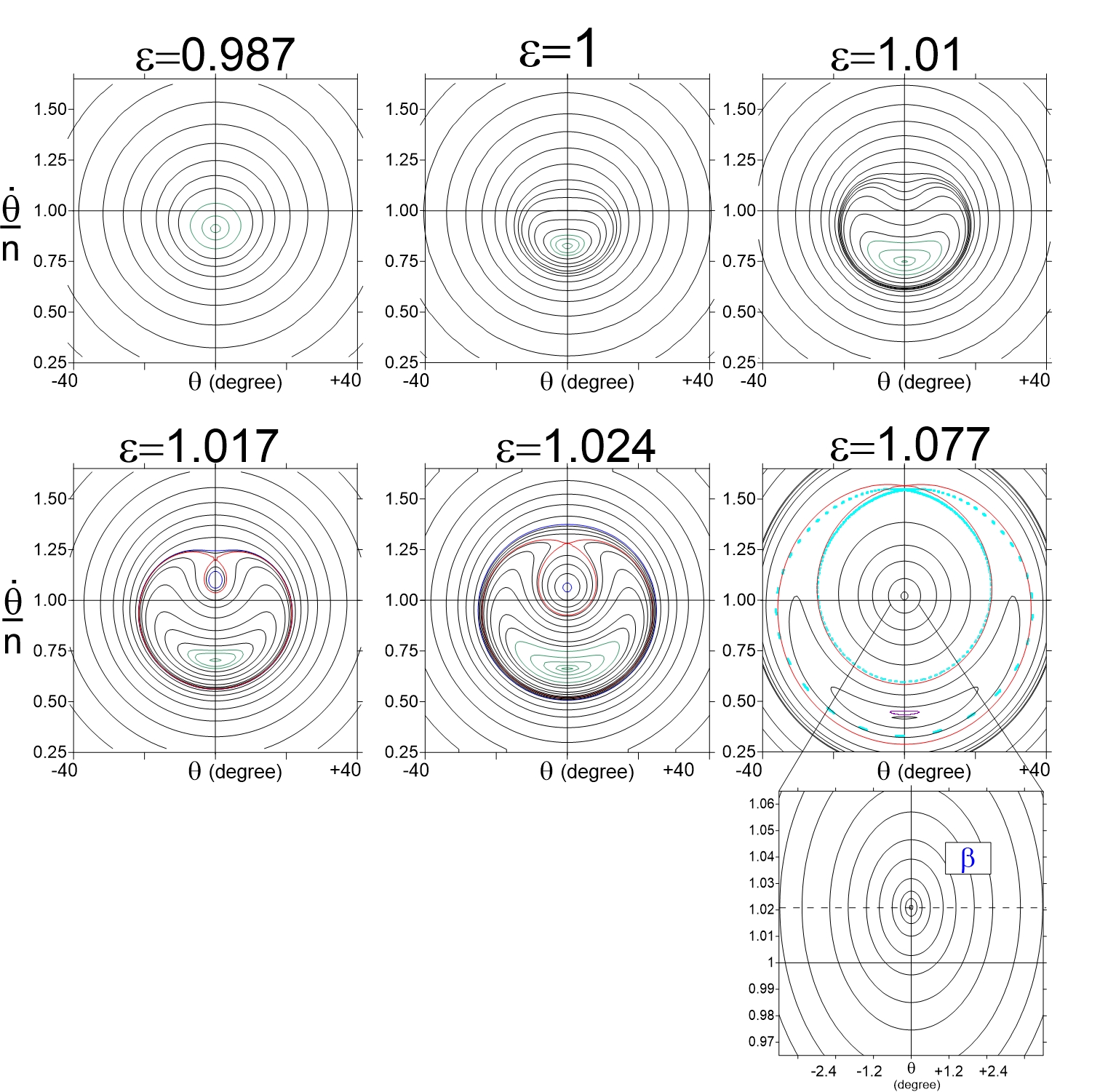}}
\caption{Contour plots of the hamiltonian (\ref{26}) for several values of the non-sphericity parameter $\epsilon$. The orbital eccentricity is $0.0014$ and the range of values of $\epsilon$ are compatible with the values of Methone, $\epsilon\sim1.077$ (the same utilized in Fig. \ref{1}). Green and blue lines correspond to level curves within the domains of the $\alpha$ and $\beta$-regimes, respectively. Red curves are the separatrix of the $\beta$-regime. Right-bottom plot is a detailed close of the fixed point associated with the $\beta$-regime. The dashed line indicates the forced component. The two curves given in blue and purple superposed to the level curves in the plot $\epsilon=1.077$ are the surfaces of section of the hamiltonian (\ref{1}).}
\label{<2>}
\end{figure}

\newpage\begin{figure}
\centerline{\includegraphics[width=35pc]{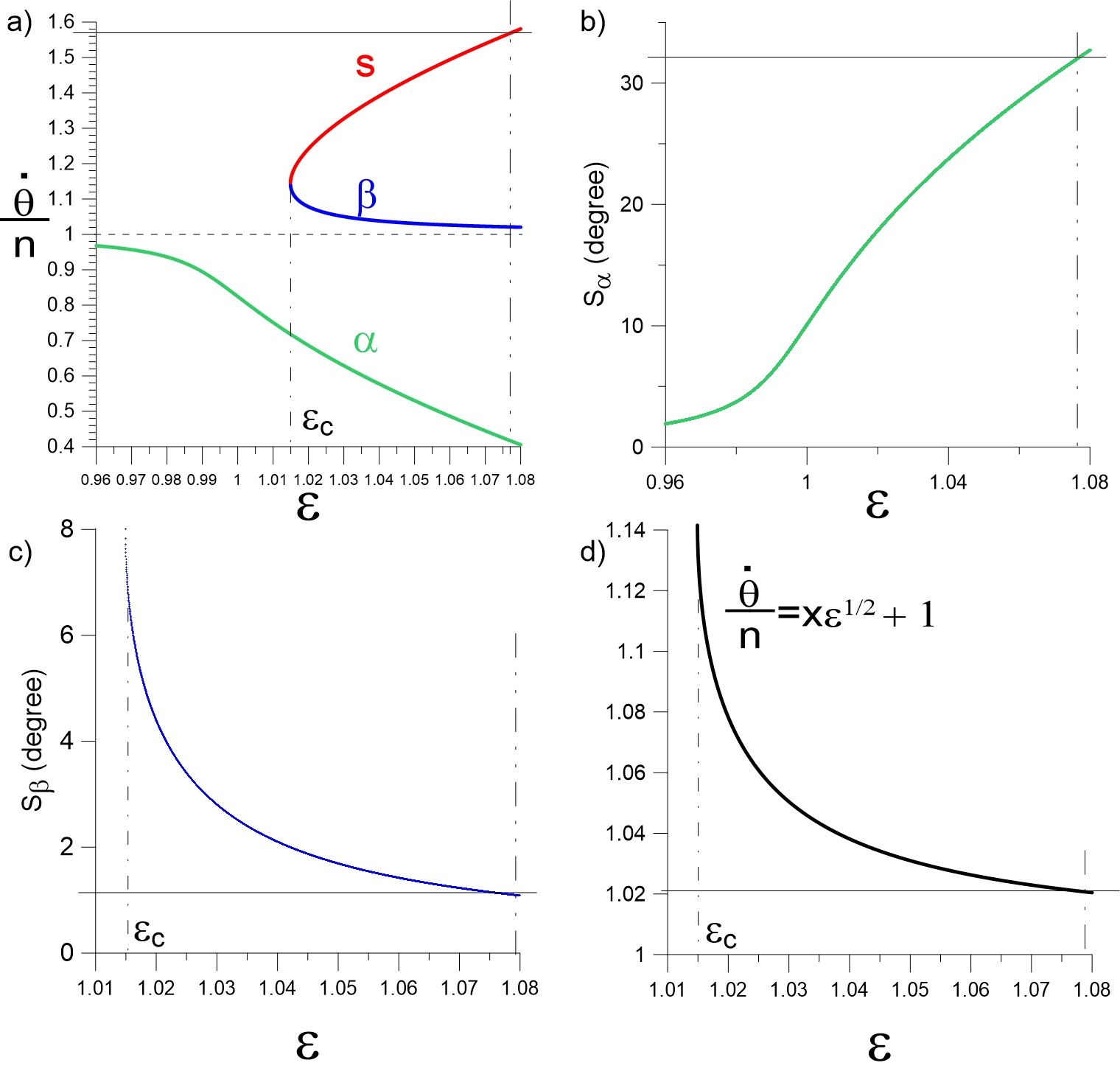}}
\caption{\textbf{(a)} The real roots of Equation (\ref{31}) as function of $\epsilon$. Red, blue and green curves are the loci of the fixed points of the separatrix ($s$), the $\beta$ and the $\alpha$-regimes, respectively. Dashed horizontal, dashed-dot vertical, and dashed-dot-dot vertical lines indicate, respectively, the straight $\frac{\dot\theta}{n}=1$, the critical value $\epsilon_c=1.0149$ where the $\beta$-regime rises, and the current value for Methone ($\epsilon=1.077$). Horizontal lines give the values at the corresponding y-axes. \textbf{(b)} Amplitude of $\phi=\theta(t)-nt$ close to fixed point of the $\alpha$-regime. \textbf{(c)} The same as (a) in the case of $\beta$-regime. \textbf{(d)} The curve of $\frac{\dot\theta}{n}=x\sqrt{\epsilon}+1$ in the case of the root $x_2$.}
\label{<3>}
\end{figure}

\subsection{The forcing of the quasi-synchronous $\beta$-regime}

The loci of the equilibria centers (or the fixed points in the full problem) of the $\alpha$ and $\beta$-regimes in the phase space can be obtained analytically from equations of motion:

\begin{eqnarray}
        \frac{dx}{dt}&=&+\frac{\partial H}{\partial y}=-\frac{1}{4}(x^2+y^2)y + \delta y + 2\gamma_1yx      \label{29} \\
        \frac{dy}{dt}&=&-\frac{\partial H}{\partial x}=+\frac{1}{4}(x^2+y^2)x - \delta x - 3\gamma_1x^2 - \gamma_1y^2 +\gamma_2,      \label{30}
\end{eqnarray}
where $\gamma_1=\frac{3}{8}e\sqrt{\epsilon}$, $\gamma_2=e\sqrt{\epsilon^3}$.

The equilibrium condition $(\frac{dx}{dt},\frac{dy}{dt})=(0,0)$ leads to a system of non-linear algebraic equations, and equilibria solutions in the x-axis where $y=0$ must satisfy the equation:

\begin{equation}
\frac{x^3}{4} - 3\gamma_1x^2 - \delta x +\gamma_2=0. \label{31}
\end{equation}

The three solutions of Equation (\ref{31}) are real only for values of $\epsilon$ larger than $\epsilon_c$. For $\epsilon<\epsilon_c$, only one root is real and there is only one fixed point in the phase space, namely, that one related to the $\alpha$-regime. At $\epsilon=\epsilon_c$ occurs the rising of the $\beta$-regime and the unstable point. Considering the current $\epsilon=1.077$ of Methone, the solution of Equation (\ref{31}) is $(x_1,x_2,x_3)=(-0.5616685,0.02034028, 0.5478663)$. Substituting these roots into Equation (\ref{28}) we obtain $\frac{\dot\theta}{n}=0.4171$ ($x_1$: $\alpha$-regime); $\frac{\dot\theta}{n}=1.0211$ ($x_2$: $\beta$-regime); and $\frac{\dot\theta}{n}=1.5686$ ($x_3$: separatrix). The three values agree with those obtained after inspection of the level curves given Fig. \ref{2}. Fig. 3a) shows the real solutions calculated in this way in the range of $0.96<\epsilon<1.08$. We can see that the bifurcation occurs at $\epsilon_c\sim1.0149$. Note also the quasi-symmetry of the roots $x_1$ and $x_3$.

The fixed point of the $\beta$-regime is not centered at the origin as it is indicated by the dashed line in Fig. \ref{2}, bottom-right. The forced component calculated above deviates from the unit such that $\frac{\dot\theta}{n}=1.0211$. More generally, note that in the case of non-circular orbits, $x=0$ is never a solution of Equation (\ref{31}), so that from Equation (\ref{28}) the forcing is never null in this case.

The forced component in the SOS can also be explained by the classical linear theory of the average solution of the spin-orbit dynamics within the synchronous resonance. The forced harmonic oscillator-analogue model gives $\frac{\dot\theta}{n}=1+A$ where the amplitude of the forced component $A$ is given by:

\begin{eqnarray}
A=\frac{2\omega^2e}{\omega^2-n^2}=\frac{2e}{1-\epsilon^{-2}}, \label{32}
\end{eqnarray}
(Callegari and Rodr\'{i}guez 2013; see also equation 5.123 in Murray and Dermott 1999). From Equation (\ref{32}) we obtain $\frac{\dot\theta}{n}=1.0203$, showing good agreement with the value obtained above.


\subsection{Amplitude of the physical libration}

We can obtain an approximate expression of the amplitude of libration of the angle $\phi=\theta(t)-nt$ from equation (24) in Wisdom (2004):

\begin{eqnarray}
S=\frac{1}{2}\left( 8\frac{\hat{J}}{\epsilon}+4\frac{\hat{J}^2}{\epsilon^2} \right)^{1/2}.\label{99}
\end{eqnarray}

The next step is to calculate the amplitude of libration of $\phi$ as a function of $\epsilon$. For this task, we need to evaluate the action $\hat{J}$ in the equation above, and following Wisdom's method, we utilize the stable equilibria points of the hamiltonian.
At equilibrium  $x=\sqrt{2\hat{J}}$, so that the action of the stable equilibrium is $\hat{J}=\frac{x^2}{2}$, where $x$ is the corresponding solution of Equation (\ref{31}).

The plot of $S$ in the case of the secondary resonance (denoted by $S_{\alpha}$) as function of $\epsilon$ is given in Fig. 3b). The amplitude of $\phi=\theta(t)-nt$ increases to values close to the current one indicated by vertical and horizontal lines. There is good agreement when compared to the fixed point in Fig. 1c) and the equilibria points given in Fig. 2.

In Fig. 3(c) it is shown the numerical value of $S$ is the case of the $\beta$-regime (denoted by $S_{\beta}$). In the case the amplitude of $\phi=\theta(t)-nt$ it is always smaller than that case of the $\alpha$-regime. Note that $S_{\beta}$ decreases as $\epsilon$ increases. This is related to the fact the forced component is larger for values of $\epsilon$ closer to the $\epsilon_c$ as seen in Fig. 3d), where $\frac{\dot\theta}{n}=x\sqrt{\epsilon}+1$ is plotted in the case of the root $x_2$. Note the good agreement with the fixed points given in Fig. 2.

The amplitude of variation of the $\phi=\theta(t)-nt$ in the $\alpha$-regime is much larger than in the case of the $\beta$-regime. The current rotation state of the regular satellites of the Solar System is the synchronous equilibrium (e.g. Peale 1977). It is well known that this final stage of rotation has been reached after a long-term evolution due to planet-satellite tidal interactions (see Wisdom 2004, Ferraz-Mello 2013, and references therein). The corresponding amplitudes of the physical libration are always very small; Enceladus has one of the largest values $\sim0.12$ degree. Therefore, the $\beta$-regime is the probable final state of rotation of these sorts of bodies.

In the case of small satellites, the existence of secondary resonances due to their irregular shapes leads to the question of the possibility of high-amplitude libration on $\alpha$-regime. For instance, in the case of Saturnian small satellites, the most recent observations and analyses give small amplitudes of libration (Thomas and Helfenstein 2020, Rambaux et al. 2022). According to Melnikov and Schevchenko (2022), the amplitude of libration of Amalthea (Jupiter) is $<5$ degree and would be $\sim30$ degree if the $\alpha$-regime had been attained (Pashkevich et al. 2021). Therefore, based on observations, it is probable that, in despite of their large $\epsilon$, all small satellites of Saturn and Jupiter also currently rotate in the equilibrium site associated with the $\beta$-regime.\\

\begin{figure}[h]
\centerline{\includegraphics[width=40pc]{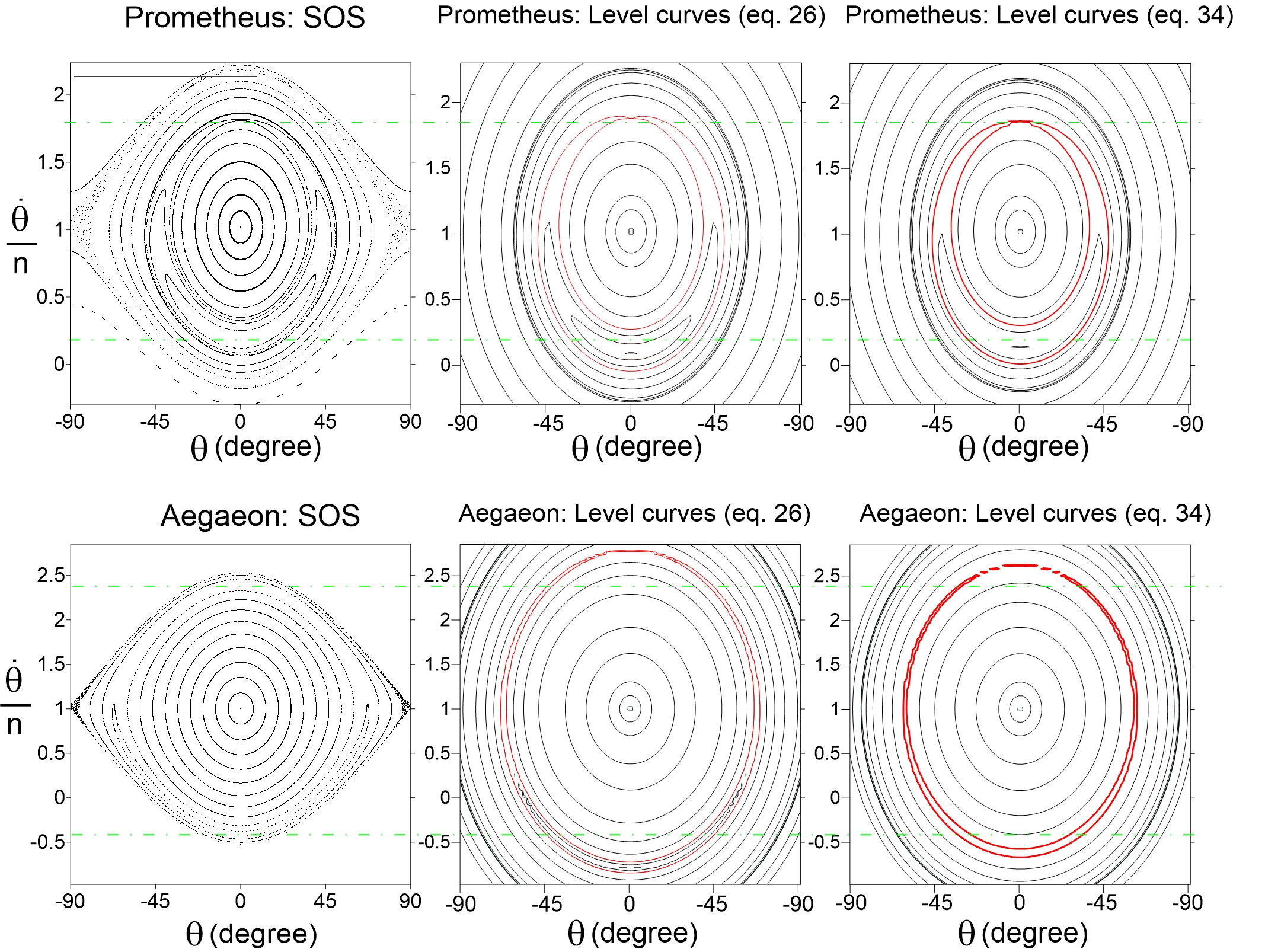}}
\caption{The surfaces of section (SOS; first column), and level curves of the hamiltonian (\ref{26}) (second column), and Equation (\ref{33}) (third column) in the cases of Aegaeon and Prometheus (top and bottom, respectively). The separatrix of the $\alpha$-regime is shown in red. Dashed-dot-horizontal green lines indicate the equilibria points obtained from hamiltonian and compared to the SOS corresponding fixed points.}
\label{<5>}
\end{figure}

\section{Additional application: Aegaeon and Prometheus.}

Fig. \ref{4} shows the rotational phase space around the synchronous regime in the cases of Aegaeon ($\epsilon\sim1.5236$), and Prometheus ($\epsilon\sim1.1727$). As pointed out in Section 1, both rotational phase spaces display the $\alpha$ and $\beta$-regimes. Inspection of the left and middle panels shows us that the position of the three equilibria points of the $\alpha$ and $\beta$-regimes obtained from level curves (middle) agree with surfaces of section in the case of Prometheus, but the same is not true in the case of Aegaeon\footnote{Note that our model does not allow to obtain the loci of the separatrix of the synchronous regime since it is valid, by construction, only in the interior of the resonance.}.

To improve the model we include terms of order higher than $J^2$ in the action so that the hamiltonian (\ref{26}) becomes:

\begin{eqnarray}
H(x,y)=\frac{\delta}{2}(x^2+y^2)-\frac{1}{16}(x^2+y^2)^2+\frac{3}{8}e\sqrt{\epsilon}(x^2+y^2)x-e\sqrt{\epsilon^3}x\nonumber\\
-\frac{1}{128\epsilon}(x^2+y^2)^3+\frac{17}{96\sqrt{\epsilon}}ex(x^2+y^2)^2\label{33}.
\end{eqnarray}
The last two terms in Equation (\ref{33}) are of the order $O(J^3)$ and $O(J^{5/2})$, respectively. The term $\frac{1}{128\epsilon}(x^2+y^2)^3$ rises from $H_0''$ (the last term at the right in the Equation (\ref{10})). The second one comes from $H_1''$ (Equation (\ref{20})) after considering the terms of order $J^{5/2}$ at the right in Equations (\ref{12}) and (\ref{13}).\\

Inspection of Figure 4, bottom, shows a significant improvement in the case of Aaegeon. As discussed in Lei (2023), and first shown in Gkolias et al. (2019), in the specific case of the 1/1 secondary resonance, the analytical and numerical mapping of the associated fixed points diverge for values of $\epsilon>1.2$.

\section{Conclusions}

This work shows the results of the dynamics of rotation of out-of-round close-in small Saturnian satellites Methone and Aegaeon discovered by the Cassini spacecraft. Their shapes and physical parameters have been updated by Thomas and Helfenstein (2020). As pointed out in Section 1, the main problem we are considering in this work is related to the nature of the $\alpha$ and $\beta$-regimes of motion in the phase space of rotation of these satellites already detected for Amalthea, Metis, and Prometheus (and probably others - see Melnikov and Shevchenko 2022, and references therein).

We first conduct our investigation with the well-known model of rotation developed by Goldreich and Peale (1966) and applied by many authors in distinct sorts of problems through decades (e.g. Wisdom et al. 1984, Wisdom 2004, Callegari and Rodr\'iguez 2013). The exact equations of the 1.5 degrees of freedom model (obtained from Equation (\ref{1})) are solved numerically with the integrator provided by Everhart (1985), resulting in accurate rotational trajectories able to identify the main properties of the phase space with the surface section technique. In the case of Methone, the $\beta$-regime is located close to the origin and corresponds to the ``classical'' fixed point associated with the synchronous rotation motion where the amplitude of oscillation of the angle $\Psi=\theta-f$ is small at the fixed point ($\sim 1$ degree). Its center is slightly forced by an amount $\sim0.02$ such that $\frac{\dot\theta}{n}\sim1.02$, a value which can be confirmed by linear theory. The $\alpha$-regime engulfs the $\beta$-regime and it is located far from origin at $\frac{\dot\theta}{n}\sim0.45$ such that $\frac{\dot\theta}{n}$ oscillates around 1. The amplitude of variation of $\Psi$ is large at the fixed point of the $\alpha$-regime ($\sim 30$ degree). The numerical results of the amplitude of oscillations agree with analytical estimative (Section 3.5).

To understand the complexity of such rotation phase space of Methone and Aegaeon, we apply the model developed by Wisdom (2004) to study the rotation of Enceladus. Thus, we follow exactly his steps and reproduce his results; having completed this task, we could apply the methodology to our case. Equation (\ref{18}) generalizes Wisdom's theory and it is ready to write the models for 1/1 and 3/1 secondary resonances up to $J^3$ in the action. The terms in the expansion involving the 1/1 resonance between the frequency of physical libration $\Psi$ and the mean motion $n$ are responsible for the co-existence of the $\alpha$ and $\beta$-regimes in the rotation of the current Methone and Aegaeon's phase spaces. Therefore, we have shown the existence of a secondary resonance within the synchronism, completing our initial goals.

Figures \ref{2} and \ref{3} show the dependence of the domains of the $\alpha$ and $\beta$-regimes with the shape parameter of Methone. The $\beta$-regime occurs at $\epsilon=\epsilon_c\sim1.0149$ such that for $\epsilon<\epsilon_c$ only the $\alpha$-regime exists. While there is good agreement between numerical and analytical calculations up to $J^2$ in the case of Methone ($\epsilon\sim1.077$), the same is not true for Aegaeon ($\epsilon\sim1.5236$), where the model is useful only when higher order terms are included in the hamiltonian since $\epsilon>1.2$, a superior limit for convergence of analytical estimative (Gkolias et al. 2019), as it is shown in Figure 4.

We revisit the previous works on the rotation of Prometheus, Amalthea, and Metis given the current theory. The case of Metis is similar to that Methone since $\epsilon\sim1.0742$ (see Table 1). The case of Amalthea is similar to that of Prometheus since $\epsilon\sim1.214$ (see Table 1).

As we pointed out above, trajectories within the $\alpha$-regime of motion are forced and suffer large oscillations even in the case of initial conditions located over the fixed point. Being that the current rotational states of Saturnian close-in small satellites are probably synchronous (e.g. Thomas and Helfenstein 2020), our results can be taken into account in evolutionary studies of the rotation of these bodies. Tidal models (e.g. Ferraz-Mello et al. 2008, Ferraz-Mello 2013) can be applied to estimate the final destiny of the rotation of such small bodies and also the role of the secondary resonances on its thermal emission (Wisdom 2004).

\vspace{1cm}

\textbf{ACKNOWLEDGEMENTS.}

The São Paulo Research Foundation (FAPESP) (process 2020/06807-7).\\


The initial results of this work has been presented at the `XXI Brazilian Colloquium on Orbital Dynamics' (INPE, from December 12 to 16, 2022); and at the `New Frontiers of Celestial Mechanics: theory and applications' (Department of Mathematics Tullio Levi-Civita, University of Padua, from February 15 to 17, 2023).

\newpage
\section*{Appendix: Dependence of the phase space of rotation of Methone with the orbital eccentricity.}

Consider two ranges of eccentricity: $A=[0.0005,0.005]$ and $B=[0.005,0.02]$.

Fig. 5  shows levels of the hamiltonian (26) for three values of eccentricity in the interval $A$. The domains of the $\alpha$ and $\beta$-regimes remain almost unchanged for eccentricities closer to the current one. For $e$ slightly larger than the current one ($e=0.005)$, the domain of the $\alpha$-regime (the secondary resonance) is enlarged, while the domain of the $\beta$-regime slightly decreases. This result is confirmed with the surface of section (SOS). Note that inspection of the two plots of bottom figures shows us that no chaotic variation is shown around the separatrix of the secondary resonance (red curve).

Discussions given above on the dependence of the $\alpha$ and $\beta$-regimes with the orbital eccentricity are restricted within the domains of the synchronous regime. To understand the role of larger orbital eccentricity in the global structure of the phase space we must consider the volume encompassing the whole synchronous resonance. Fig. 6 shows SOS for some values of eccentricity given in the interval $B$. The chaotic layer of the synchronous resonance increases as a function of the eccentricity. The domain of the $\alpha$-regime remains almost unchanged while the $\beta$-resonance is diminished. For eccentricities $e\sim0.018$, the $\beta$-regime becomes chaotic, and for $e\sim0.02$ it is overlapped with the synchronous chaotic layer.

 \begin{figure}
 \centering
 \includegraphics[width=16cm]{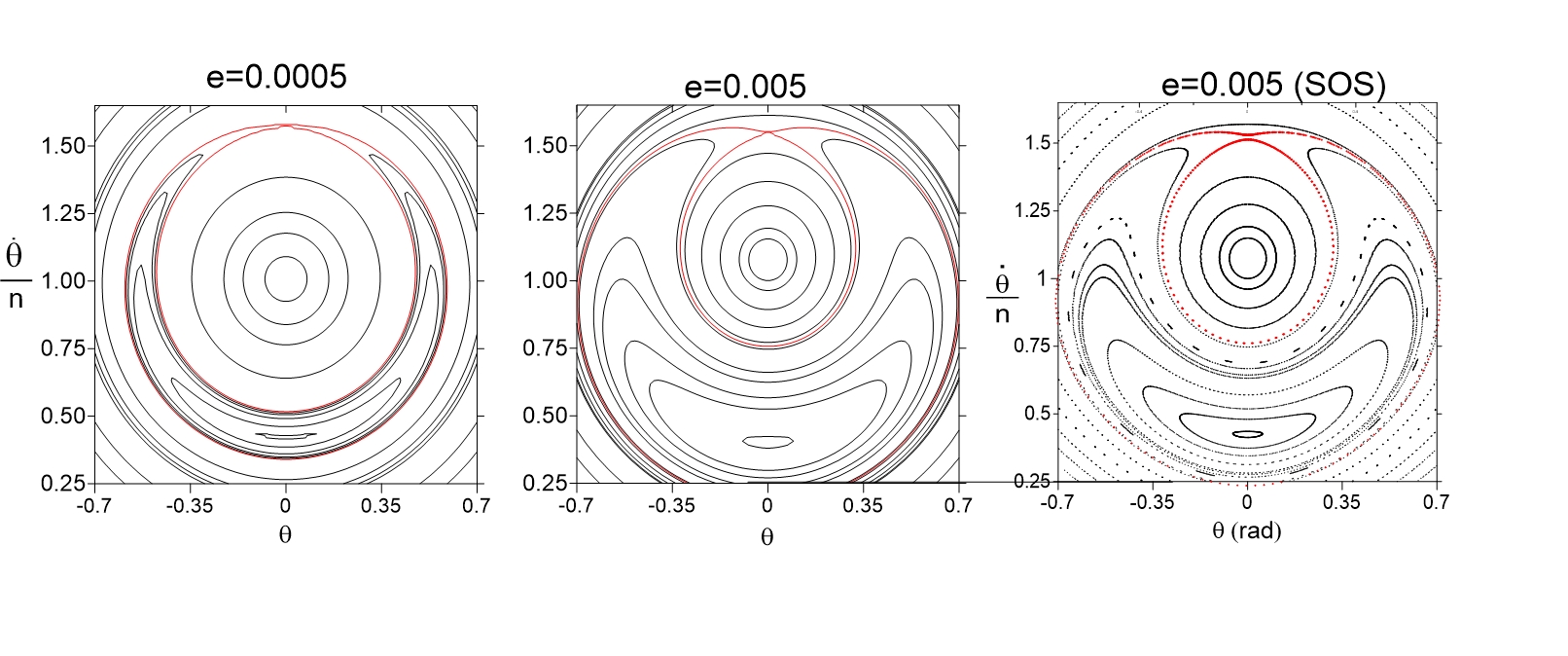}
  \caption{Level curves of hamiltonian (26) for $\epsilon=1.077$ for different values of eccentricity. The plot given at right shows the surfaces of sections of solutions of hamiltonian (1) in the case $\epsilon=1.077$, $e=0.005$.}
  \label{<F3>}
\end{figure}

 \begin{figure}
 \centering
 \includegraphics[width=16cm]{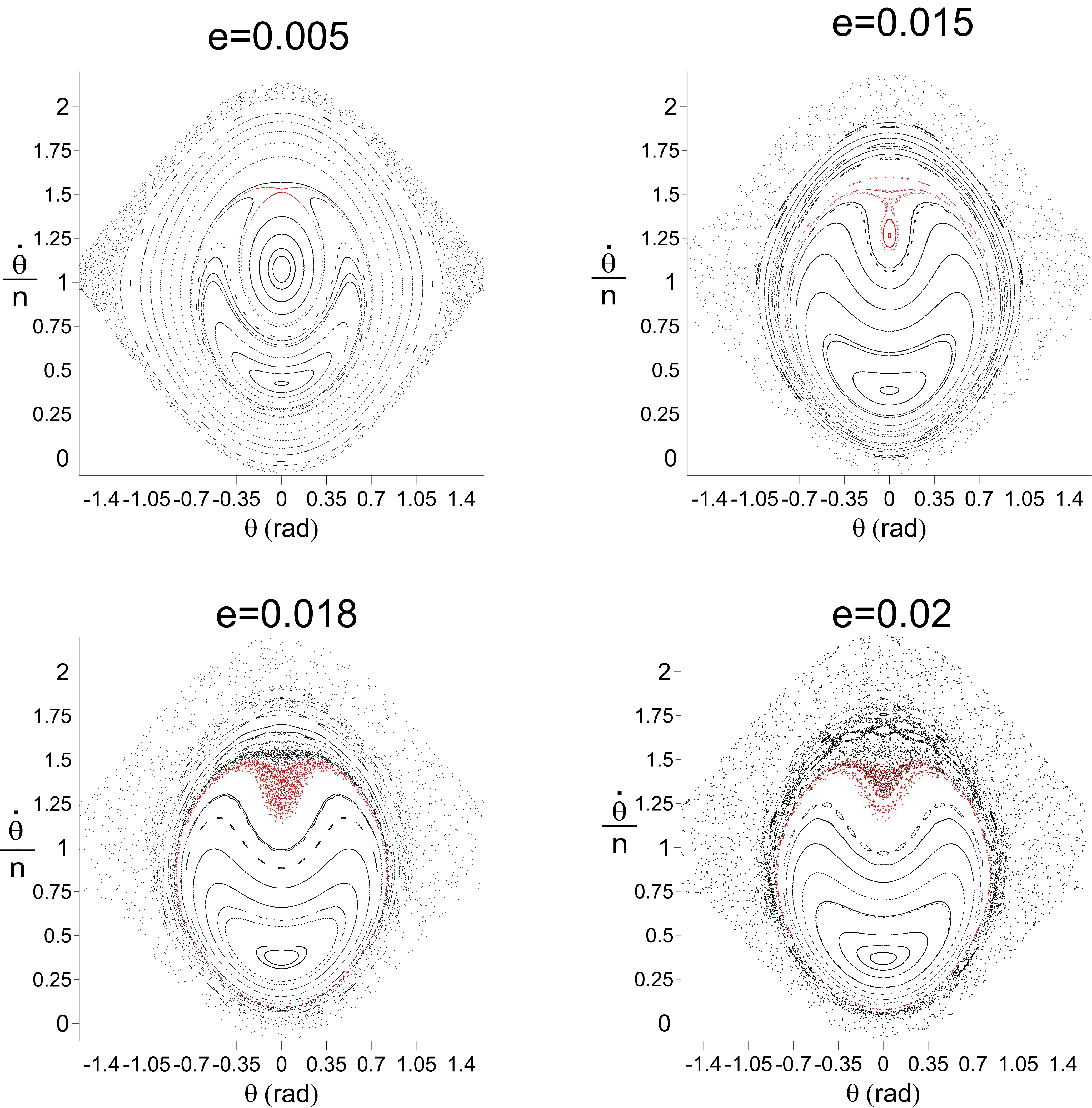}
  \caption{Surfaces of sections of solutions of hamiltonian (1) for $\epsilon=1.077$ and different values of eccentricity indicated at the top of the plots.}
  \label{<F3>}
\end{figure}

\end{document}